\documentclass[journal=jacsat,manuscript=article]{achemso}

\usepackage[version=3]{mhchem} 

\usepackage[utf8]{inputenc} 
\usepackage[T1]{fontenc}    
\usepackage{hyperref}       
\usepackage{url}            
\usepackage{booktabs}       
\usepackage{amsfonts}       
\usepackage{amsmath}
\usepackage{nicefrac}       
\usepackage{microtype}      
\usepackage{array} 
\usepackage{varwidth}
\usepackage{graphicx}
\usepackage{multirow}
\usepackage{subfig}
\usepackage{amssymb}
\usepackage{array}
\usepackage[section]{placeins}
\usepackage{afterpage}

\author{Ahmadreza Ghanbarpour}
\affiliation{Department of Medicinal Chemistry and Molecular Pharmacology, College of Pharmacy, Purdue University, 575 Stadium Mall Drive, West Lafayette, Indiana 47906, United States}

\author{Markus A. Lill}
\affiliation{Department of Pharmaceutical Sciences, University of Basel, Klingelbergstrasse 50, 4056 Basel, Switzerland}
\email{markus.lill@unibas.ch}
\phone{++41 61 2076135}
\alsoaffiliation{Department of Medicinal Chemistry and Molecular Pharmacology, College of Pharmacy, Purdue University, 575 Stadium Mall Drive, West Lafayette, Indiana 47906, United States}

\title[An \textsf{achemso} demo]
  {Seq2Mol: Automatic design of de novo molecules conditioned by the target protein sequences through deep neural networks}

\abbreviations{IR,NMR,UV}

\begin{document}






\begin{abstract}
De novo design of molecules has recently enjoyed the power of generative deep neural networks. Current approaches aim to generate molecules either resembling the properties of the molecules of the training set or molecules that are optimized with respect to specific physicochemical properties.
None of the methods generates molecules specific to a target protein.
In the approach presented here, we introduce a method which is conditioned on the protein target sequence to generate de novo molecules that are relevant to the target. We use an implementation adapted from Google's "Show and Tell" image caption generation method, to generate SMILES strings of molecules from protein sequence embeddings generated by a deep  bi-directional language model ELMo. ELMo is used to generate contextualized embedding vectors of the protein sequence. Using reinforcement learning, the trained model is further optimized through augmented episodic likelihood to increase the diversity of the generated compounds compared to the training set. We used the model to generate compounds for two major drug target families, i.e. for GPCRs and Tyrosine Kinase targets. The model generated compounds which are structurally different form the training set, while also being more similar to compounds known to bind to the two families of drug targets compared to a random set of molecules. The compounds further display reasonable synthesizability and drug-likeness scores.

\end{abstract}

\section{Introduction}
De novo design of molecules is an important approach for the discovery and development of drugs. 
In recent years, artificial intelligence-based methods, in particular deep learning-based methods, have been employed to facilitate this process and open new possibilities for the design of new molecules. 
Many generative types of architectures have been used for the task of de novo molecule generation, such as autoencoder-based models, generative adversarial neural networks (GANs), recurrent neural networks (RNNs) and models combined with reinforcement learning \citep{generativemodelsreview}.
Those approaches have been focused on the generation of compounds based on a pool of training compounds and have been mostly focused on engineering molecules with specific physicochemical properties. Whereas most approaches display inherent structural similarity of the generated molecules to the original “seeds”, some newer approaches aim to increase the diversity of generate molecules compared to the training set \citep{organic}. 
In the context of drug design, current approaches aim to generate novel compounds that resemble features (e.g. physicochemical properties or pharmacophore features) from already known binders to a specific target. A major drawback to this approach is that in many instances, e.g. when a target is newly discovered, there is no known ligand for the specific target. Therefore, no known pharmacophore elements  or physicochemical features of known compounds can guide the de novo molecule generation process. 
Even if compounds for a target are known they are often limited in number, making a target specific training without overfitting unlikely. 
Therefore, there is a need for methods to be able to condition the generative process on the biochemical features of the target. 

Our method for the generation of de novo compounds is conditioned on the sequence of a target protein. The concept is based on the “Show and Tell” image captioning method developed by Vinyals et al \citep{showandtell}. 
The question is, what do image captioning and de novo molecule design have in common? To answer this question, we take a closer look of how the "Show and Tell" image captioning method functions. In image captioning, the task is to generate a sentence with the following two essential properties: First, it should be relevant to the image, and second it should be grammatically correct and meaningful. In image captioning, essential features of a given image are extracted, e.g. using a neural network. The embedding vector is then used to generate a meaningful sentence which describes the image. 
Similarly, in de novo compound generation essential features of the target protein are extracted ('image features') and molecules are generated in form of SMILES strings ('caption').
The target protein is represented by its sequence. A neural network learns embeddings that should represent the features of the target essential for ligand binding. These features are linked to their corresponding 'caption', that is, the characters of the SMILES strings representing the chemical compound. The caption generator network also tries to learn to generate valid and meaningful sentences with the correct grammar, which is also critical in the task of SMILES generation to obtain chemically valid molecules.

In this work, we use protein sequence embeddings generated using a pre-trained bi-directional language model ELMo, and use an LSTM model combined with reinforcement learning to generate SMILES strings of de novo compounds for two important target families: GPCR and Tyrosine Kinases. Protein-ligand datasets published in BindingDB are used for training and validation of the model \citep{bindingdb}.

\section{Methods}
\subsection{Datasets}
We used the dataset from the work by Karimi et al \citep{deepaffinity} which originates from BindingDB \citep{bindingdb}. The original dataset which contains all IC$_{50}$-labeled ligand-target pairs from BindingDB was reduced to protein-ligand pairs with IC$_{50}$-values of less than 1 mM. The data was directly taken from “BindingDBAll2018m8.tsv.zip” file provided by BindingDB, which contains protein sequences as well as their corresponding 2D compounds structure in SMILES.
For correct and independent validation of the resulting model on two families of protein targets, GPCRs and Tyrosine Kinases, any protein-ligand pair matching one of those two classes was removed from the training based on records from the Uniprot database \citep{uniprot}.
Also, any compound that existed both in training set and test sets (even when bound to a different target than GPCR or Tyrosine Kinases) was removed from the test sets. 
This resulted in a training set of 127546 entries and test sets for GPCR and Tyrosine Kinases with 276 and 109 target proteins and 37749 and 25578 binding molecules, respectively.

\subsection{General workflow}
\begin{figure}[!htb]
  \centering
  \includegraphics[width=0.9\linewidth]{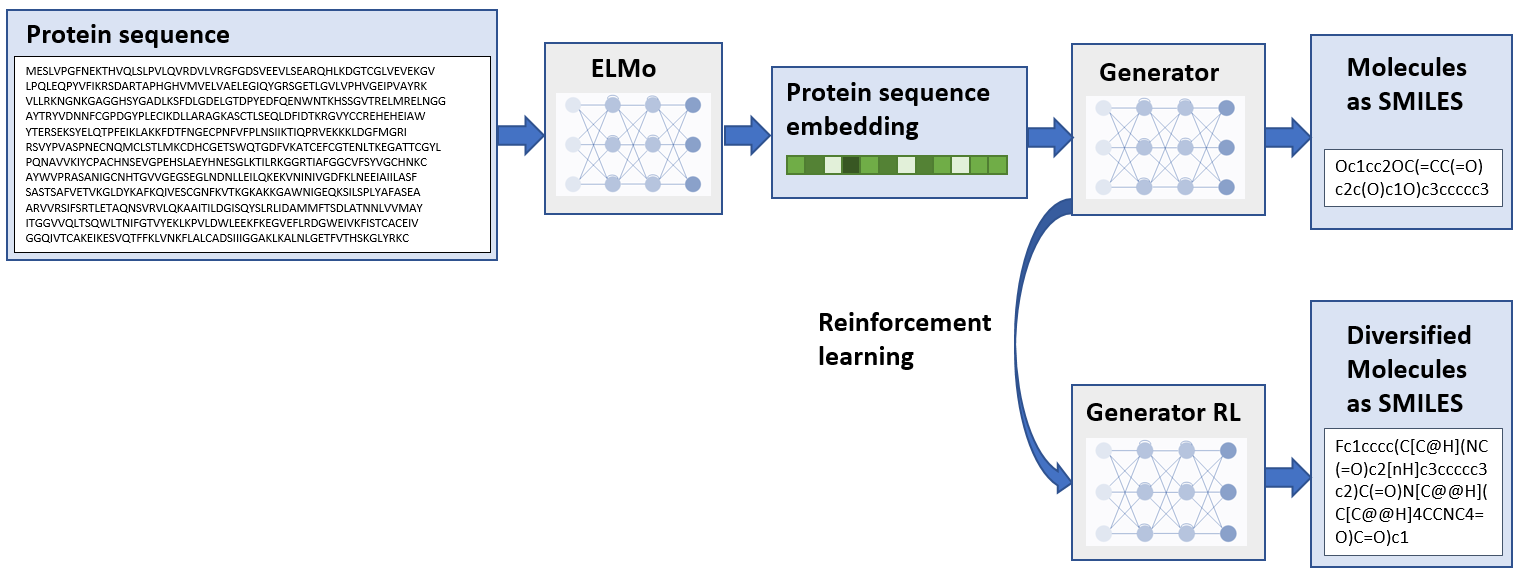}
  \caption{Overall workflow of de novo compound generation method using deep neural networks. First, sequence embeddings are generated using the network from Heinzinger et al \citep{secvec}. Then the compound generator is trained using the embeddings as input. After the initial training, the network is re-trained using a reinforcement learning scheme using the dissimilarity to the training set as reward to get more diverse compounds. }
  \label{fig_workflow}
\end{figure}

Figure \ref{fig_workflow} shows the overall workflow of the method described in this paper. Our method takes advantage of the embedding generator network developed by Heinzinger et al \citep{secvec} which generates the embeddings of protein sequences. Then the embeddings are used as inputs to the molecule generator network for the initial training. Subsequently, the model is retrained using reinforcement learning to increase the diversity and novelty of the generated compounds.

\subsection{Fingerprint generation}
We used Morgan fingerprints available in RDKit library \citep{rdkit} to analyze similarities between compounds. A radius of 4 and the bit vector length of 2048 was used to generate the fingerprints. We used Tanimoto distance (1 - Tanimoto similarity) to report the similarity in our studies, with a value of zero measuring exact identity while a value of one indicating complete dissimilarity of molecules.

\subsection{Random molecule set generation}
To measure the target specificity of the molecules generated with our neural network, those compounds were compared with randomly selected compounds. 
The latter molecules were selected from the emolecules database \url{www.emolecules.com} using the following criteria: Molecules were only selected if their log P value and molecular weight is similar to the corresponding values of known GPCR and Tyrosine Kinase binders, respectively. 
In detail, only compounds with log P values and molecular weight are selected that deviate by less than the standard deviation from the mean of the corresponding values of the known binders for the two target families. All properties were calculated using the RDKit package.
This selection process guarantees to test the model for its ability to generate target-specific ligands and not just compounds with similar physicochemical properties as known binders.

\subsection{Generation of protein sequence embeddings}
To generate embeddings of protein sequences we used the model provided by Heinzinger et al \citep{secvec}. Embedding vectors of length 1024 are generated. 
The network consists of one CNN layer and two bidirectional LSTM layers, which provide context information of the surrounding residues. For any query protein sequence, the output feature vectors of length 1024 of all three layers were summed component-wise and finally averaged over all residues of the whole protein sequence to obtain a single vector representing a protein sequence.
This sequence-embedding vector is used as input for the molecule generator network.

\subsection{Compound generation from protein sequence embeddings}

\subsubsection{Architecture of generator model}
\begin{figure}[!htb]   \centering
  \includegraphics[width=0.9\linewidth]{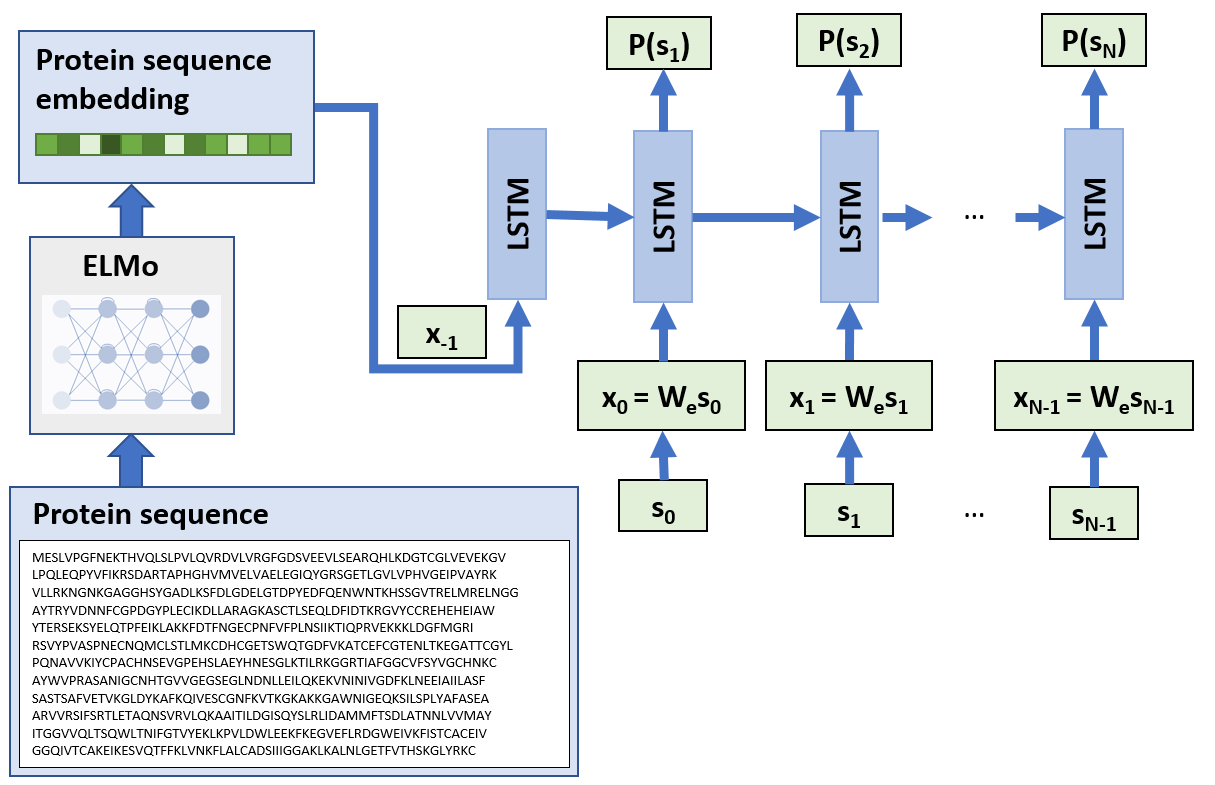}
  \caption{Molecule generator network is defined by combining LSTM model and protein sequence embedder model. The LSTM model is showed in unrolled form, where recurrent connections are shown as feed-forward connections. During training, target sequence tokens ($s_t$) are learned by maximizing $P(s_t)$, where $t$ denotes the character position in the SMILES string. Each token is passed through an embedding layer prior to LSTM.}
\label{fig_lstm}
\end{figure}

The image-captioning network architecture "Show and Tell"\citep{showandtell} based on a Keras \citep{keras} implementation was used to generate molecules in form of SMILES strings.
The model is a LSTM-based sentence generator based on given embeddings. The LSTM model's task is to predict a new SMILES character based on previously predicted SMILES characters and the protein's sequence, with probability $p\left(s_{t} | I, s_{0}, \ldots, s_{t-1}\right)$, where $I$ is the  protein's sequence embedding, and $s_t$ is the SMILES character at position $t$. 
The model is illustrated in Figure \ref{fig_lstm}. For any given position $t$ the output of the LSTM cell depends on the cell state which is the result of the current input and previous cell state at position $t-1$. Therefore, the LSTM network keeps memory of past characters. 
The protein's sequence embedding is only used for the initiation of the LSTM cell and hidden states. Due to the recurrence of LSTM networks the protein sequence, however, influences the cell state for subsequent character predictions.
\begin{equation} \label{eq1}
x_{-1} =\operatorname{ELMo}(I) 
\end{equation}
Thus the prediction of all characters is influenced by the conditioning from the protein's sequence. 

Each possible character in the SMILES string $S$ at position $t$, $s_t$, is tokenized and represented as one-hot vectors. Every sequence begins with a special character representing the start of the string at position $t=0$ and ends with an end character token (position $t=N$). Each character token passes through an embedding layer ($W_e$) prior to being input to the LSTM cell. 
\begin{equation} \label{eq1a}
x_{t} =W_{e} s_{t}, \quad t \in\{0 \ldots N-1\} 
\end{equation}

The LSTM cell predicts the probability for a character at position $t+1$ by
\begin{equation} p(s_{t+1}) =\operatorname{LSTM}\left(x_{t}\right), \quad t \in\{0 \ldots N-1\}
\end{equation}

The loss is computed by summing the negative log likelihood of each correct character token:

\begin{equation} \label{eq2}
L(I, S)=-\sum_{t=1}^{N} \log p_{t}\left(s_{t}\right)
\end{equation}

The SMILES strings of the known training compounds were tokenized in character-level and were fed to the network for training (Figure \ref{fig_lstm}). 
The network was trained for 50 epochs with batch size 512. The dimensions of the embedding layer ($W_e$) was set to 2048, and the protein embeddings were simply tiled to have the same dimensions, as the character embedding layer, i.e 2048. The dimensions of the one-hot encoding vectors were $47 \times 102$ which are the number of possible SMILES characters in the dataset and the maximum length of the SMILES string, respectively.

\subsubsection{Generation of new molecules}
\begin{figure}[!htb]
  \centering
  \includegraphics[width=0.5\linewidth]{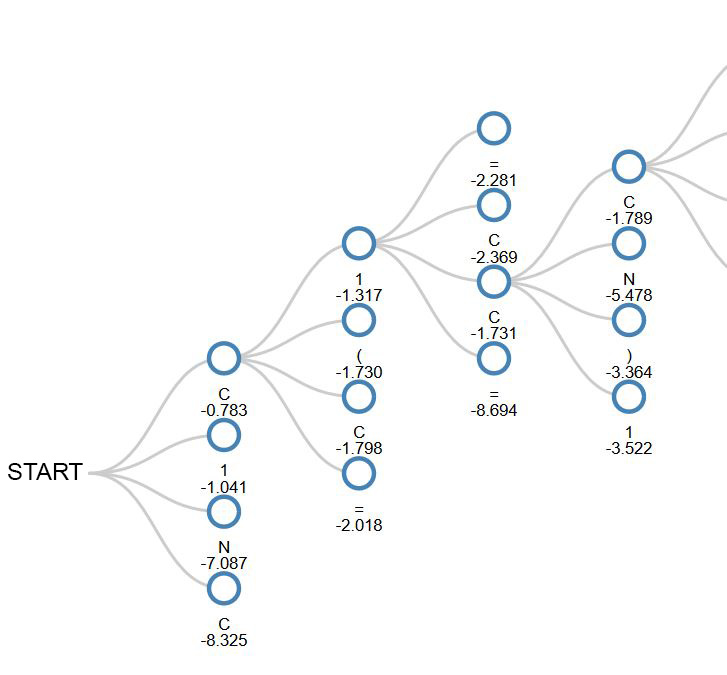}
  \caption{An example of beam search for generating new molecules. In each  step, character candidates are ranked based on scores (natural logarithm of probabilities predicted by the network). Top $k$ best candidates are considered. At each step a path is generated from one layer to the next layer forming a tree. Each path from the start token to the end token is considered a full SMILES string (molecule). For simplicity and a clearer illustration, only one path and a segment of the full tree is shown here.}
  \label{fig_beam_search}
\end{figure}

The trained model can be used to generate new molecules for a given protein sequence. 
The protein's embedding and the start token are given as initial input to the network (cf. Figure \ref{fig_lstm}). Subsequently, the LSTM network is utilized to generate the characters of the SMILES string until the end token was selected as output of the LSTM. 
To increase the diversity of the generated molecules, the LSTM network is used within the framework of a beam search. Thus, not only one character is selected at each position $t$ but the top $k$ SMILES strings at position $t$ are selected and passed on to the next LSTM iteration, keeping the best $k$ strings and so forth. (Figure \ref{fig_beam_search}). In our case, we used a value of 46 for $k$. 
This approach allows to generate a diverse set of molecules specific to the input sequence of the target protein.
Whereas selecting only one (the most probable) character at each position (greedy search) is the best choice for that specific position, it often results in sub-optimal solutions when the full string is considered \citep{showandtell}. 
In de novo compound generation, it is often desirable to have a diverse set of compounds generated for a target, rather than just one compound, especially when invalid or already known compounds are obtained by using a greedy search.

\subsubsection{Reinforcement learning}
\begin{figure}[!htb]
  \centering
  \includegraphics[width=1\linewidth]{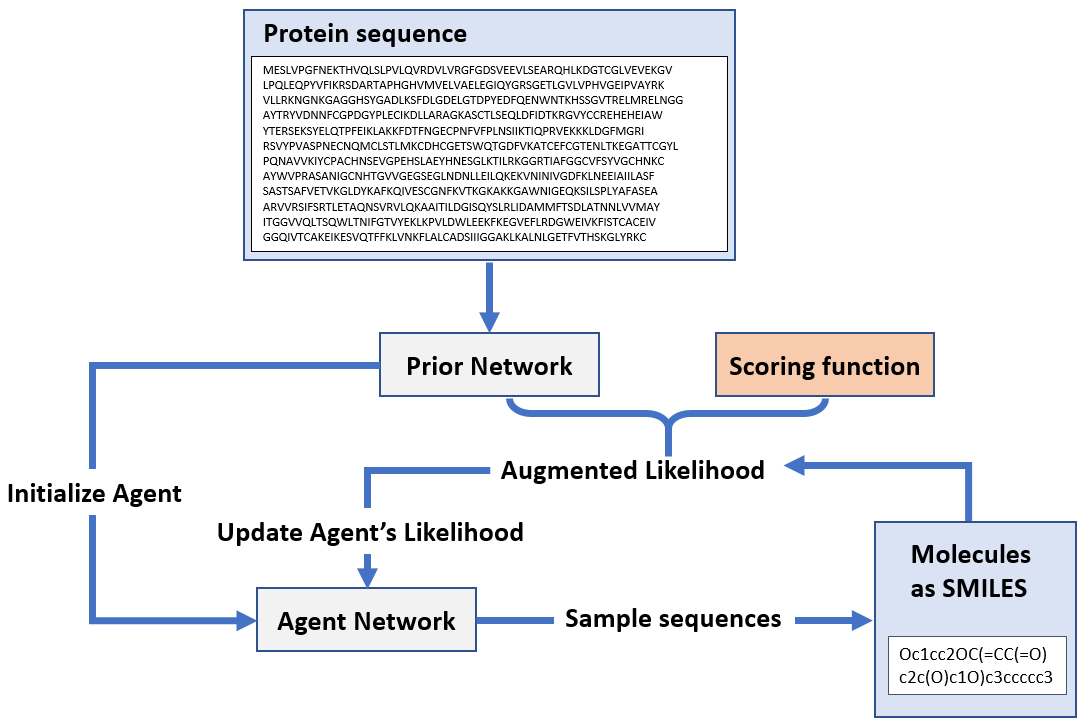}
  \caption{Encouraging diversity and novelty of generated molecules through reinforcement learning. First, the Agent network is initialized from the already trained Prior network. The Prior likelihood is then augmented by the addition of a score that measures the structural diversity of the generated compound to all training molecules. This likelihood is used to train the Agent network.}
  \label{fig_RL}
\end{figure}

To increase the novelty of the compounds when compared to the training set, a reinforcement learning procedure was employed using a concept adapted from the work by Olivecrona et al \citep{reinvent}. 
Figure \ref{fig_RL} illustrates the reinforcement learning procedure. The network that was initially trained following the procedure described in the in the previous section functions as Prior. The Agent is instantiated by a copy of the Prior network. Thus, initially, both the Prior and the Agent are identical. 
During reinforcement learning, a policy is learned by the Agent to generate compounds with desired features, here potential binders to a target protein but diverse to the initial training set. 
As described in the previous section, generated compounds are represented by SMILES strings. Those strings are generated by sampling one character at each LSTM step until the end token is reached. 
This process can be considered as a set of actions $A=a_1,a_2,...a_N$ that composes an episode of a SMILES string generation. The likelihood for a SMILES string generated by the model is 
\begin{equation}
P(A)=\prod_{t=1}^{N} \pi\left(a_{t} \mid x_{t}\right)
\end{equation}
where $\pi$ is the policy learned by the model and $x_{t}$ being the input to the LSTM at step $t$. 

To increase the likelihood the generation of SMILES strings different to the training set, a scoring function $\operatorname{\Sigma}(A)$ is added to the Prior log-likelihood 
\begin{equation}
\log P(A;\theta')_{\text{Augmented}}=\log P(A;\theta')_{\text {Prior}}+\sigma \operatorname{\Sigma}(A)
\end{equation}
where $\operatorname{\Sigma}(A)$ measures the diversity of the generated compound with respect to the training set. 
Thus, high $\log P(A;\theta')_{\text{Augmented}}$ is achieved by the generation of SMILES with high probability based on the Prior network and with diversity to the training set.
$\theta'$ are the trained weights of the Prior network.
$\sigma$ is a user-defined coefficient, which in our case was set to 60. 

Using this augmented log-likelihood the Agent's policy $\pi$ is updated from the Prior's policy $\pi_{Prior}$ to  approximate the augmented likelihood $\log{P(A;\theta')_\text{Augmented}}$. Thus, the weights $\theta$ of the Agent's network are optimized using the loss function
\begin{equation}
    L(A;\theta)=\left[\log P(A;\theta')_{\text{Augmented}}-\log P(A;\theta)_{\mathbb{A}}\right]^{2}
\end{equation}
which measured the squared difference of the current Agent's likelihood over a set of actions $A$, $\log{P(A)_\mathbb{A}}$, and the augmented likelihood.

The Agent was trained for 100 iterations with a batch size of 512. 

In our case, the goal was to increase the diversity of the generated compounds compared to the training set. 
To achieve this goal, Morgan fingerprints are computed for all training data and each generated molecule. 
The Tanimoto distances between the fingerprint of the generated molecule ($m_g$) and the set of fingerprints of all training molecules ($M_t$) are computed.
The scoring function to calculate the reward for a generated molecule is then determined by identifying the minimum Tanimoto distance ($\operatorname{T_d}$):
\begin{equation}
\operatorname{\Sigma}(A)=\min_{\forall m_t \in M_t} \operatorname{T_d}(m_g,m_t)
\end{equation}
Therefore, compounds with higher Tanimoto distance compared to the training data are rewarded.

\subsection{Benchmark}

The MOSES framework \citep{moses} was used to compare our method with other approaches to generate molecules. The MOSES framework provides pre-defined models and metrics available for comparison. The metrics and models that were used to compare the performance of our model with other models  are briefly described below.

\subsubsection{Benchmark metrics}
\paragraph{Fragment similarity.} The BRICS algorithm \citep{brics} in RDKit is used to fragment molecules and measure the cosine distance between fragment frequencies vectors:
\begin{gather*}
\mathrm{Frag}(G, R)=1-\cos \left(f_{G}, f_{R}\right)
\end{gather*}
$f_{G}$ and $f_{R}$ represent frequency vectors of generated and reference molecule, respectively. The size of the fragment vocabulary of the whole data set determines the size of the frequency vector and the elements of the vectors are the frequencies for each fragment in the molecules.

\paragraph{Scaffold similarity.} This metric measures the cosine similarity between vectors representing the scaffold of generated (G) and reference (R) molecules:
\begin{equation}
\mathrm{Scaff}(G, R)=1-\cos \left(s_{G}, s_{R}\right)
\end{equation}
The scaffolds are generated using Bemis–Murcko scaffolds algorithm \citep{bemis} implemented in RDKit.

\paragraph{Distance to the nearest neighbor.} The similarity is computed by averaging over the Tanimoto similarity value ($T$) between a molecule $m$ in the generated and reference sets. The default configurations of the MOSES framework was used to generate Morgan fingerprints for this task.
\begin{equation}
\mathrm{SNN}(G, R)=1-\frac{1}{|G|} \sum_{m_{G} \in G} \max _{m_{R} \in R} T\left(m_{G}, m_{R}\right)
\end{equation}
\paragraph{Internal diversity}
This metric measures the diversity among the molecules within the generated set.
\begin{equation}
\mathrm{IntDiv}(G)=1-\frac{1}{|G|^{2}} \sum_{m_{1}, m_{2} \in G} T\left(m_{1}, m_{2}\right)
\end{equation}

\paragraph{Other metrics} In addition to metrics above, other commonly used metrics were used to evaluate the quality of the generated molecules. The metrics used are the following:

\begin{itemize}
\renewcommand{\labelitemi}{}
\item\textbf{LogP}: The water-octanol partition coefficient, calculated using the approach of Crippen [78].

\item \textbf{Synthetic accessibility score}: \citep{sascore} A score to estimate synthetic accessibility of a molecules. Values closer to 1 indicate the compound is likely to be synthetically accessible, while values closer to 10 are expected to be difficult to synthesize.

\item \textbf{Quantitative Estimation of Drug-likeness (QED)}: A metric developed by Bickerton et al. \citep{qed} which addresses the drug-likeness of molecules based on the notion of desirability and can range between 0 to 1.
\item\textbf{Natural product-likeness score}: A measure to estimate into which of the following three categories a molecule will fall into: (1) A natural product (score between 0 and 5), (2) a synthetic product ([-5, 0]) and a drug molecule ([-3, 3]). The metric uses several substructure descriptors to determine the score \citep{natproduct}
\item\textbf{Molecular weight}: The sum of atomic weights in a molecule.
\end{itemize}
\subsubsection{Benchmark models and training}
All models used to benchmark against our model are pure ligand-generation models without consideration of any information about the target protein. The models generate novel SMILES strings based on known molecules that are also represented as SMILES. We used only the ligands (without target sequences) from our training set as training data for the benchmark models to generate novel molecules.
\paragraph{Character-level recurrent neural networks (CharRNN)\citep{charrnn}:} This model considers the SMILES as a language model and treats each SMILES character as a word.
CharRNN contains three LSTM layers and each hidden is of size 768. Dropout layer are added with dropout probability of 0.2. A Softmax function is used as the activation function of the output layer. Adam optimizer \citep{adam} is used to optimize the model's parameters using Maximum likelihood estimation (MLE). The training is done in 80 epochs with batch size equal to 64 with a learning rate set to $10^{-3}$ which is halved every 10 epochs. We used the model implemented in MOSES. 
\paragraph{Variational Autoencoder (VAE)} This model consists of two components, an encoder and a decoder. The encoder maps the input data to a lower-dimensional representation (embedding) and the decoder converts it back. 
The encoder is a bi-directional Gated Recurrent Unit (GRU) with a linear activation function. 
The decoder consists of three GRU layers with size of 512 and dropout layers with probability of 0.2. The training is done in 100 epochs with batch size of 128 by using Adam optimizer with learning rate set to $3\times10^{-4}$ to minimize the loss containing reconstruction loss  between reconstructed and input SMILES strings and Kullback-Leibler  (KL) divergence in latent space. The KL term weight is linearly increased from 0 to 1 during the training. Gradient clipping with the value set to 50 is used.
\paragraph{Adversarial Autoencoder (AAE)}
In this architecture of autoencoder the Kullback-Leibler divergence loss is no longer present. Instead, an adversarial loss is used to train the generator model in form of a discriminator network which is trained simultaneously with the autoencoder \citep{adversarialae}. The encoder and decoder are created from a 1-layer bidirectional LSTM and a 2-layer LSTM respectively, both with size of 512, and an embedding layer with a size of 128 which is shared by both. The discriminator is composed of two fully connected layers with size of 640 and 256, using Exponential Linear Unit (ELU) activation function. The model uses Adam \citep{adam} optimizer trained for 120 epochs with batch size 512. Learning rate is halved every 20 epochs with the initial value of $10^-3$.

\section{Results and Discussion}
\subsection{Protein's sequence embeddings for ligand generation}
The type of ligand that can bind to a target protein depends on the topology and physicochemical properties of the binding site of the protein.
The form and properties of the binding site is determined by the structure of the protein, which is dictated by the sequence of the protein.
As there exist only 20 natural amino acids but many thousands of structurally diverse protein structures, the sequence of the amino acids gives the structural "meaning" to the protein object. This is similar to Natural Language Processing (NLP) where an enormous number of different sentences are constructed from a smaller library of words.

\begin{figure}[!htb]
  \centering
  \includegraphics{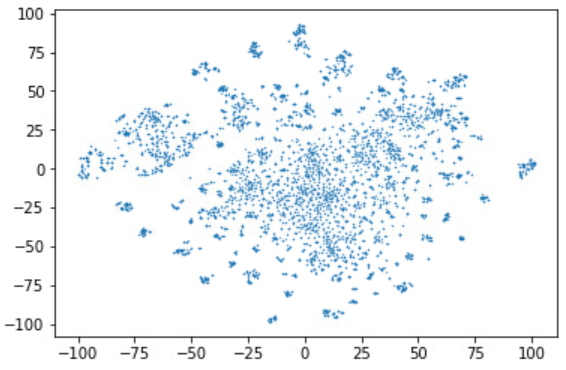}
  \caption{Sequence embeddings of protein targets in our data set generated by SeqVec, visualized using T-SNE. }
  \label{fig_sequence_embeddings}
\end{figure}

It is common to use word embeddings in the field of Natural Language Processing (NLP). Similar words can have different contexts appearing in different sentences, represented by different embeddings. In the same way, models used in NLP such as Embeddings from  Language  Model (ELMo) can be used to generate contextualized embeddings of the sequences.
Heinzinger et al. trained such a model on the Uniref50 dataset. Using a fixed model for sequence embedding that has been already trained on such a large dataset (33 M sequences) provided better performance in our study compared to a network that learns the sequence embeddings in parallel to the training of a generator network for molecules. One major reason for this observance is that the existing protein-ligand binding datasets contain a much smaller number of protein sequences compared to Uniref50. 
This approach was also taken in the work "Show and Tell", by using embeddings generated from VGG16 trained on ImageNet, one of the largest image datasets available. The actual data set, that the "Show and Tell" model was trained on, was much smaller. 
Figure \ref{fig_sequence_embeddings} shows the separation of embedding vectors for all protein sequences in our data set as generated by the SeqVec model. The embedding vectors are projected on a reduced 2D representation using T-SNE available in scikit-learn library \citep{scikit-learn}. 

\begin{figure}[!htb]
  \centering
  \subfloat[Similarity of the generated molecules (Gen) for Tyr kinases to the test set (Test) that contains Tyr kinase ligands. Higher similarity was observed compared to the similarity between training (Train) and test set or randomly selected molecules (Random) to test set.]{\includegraphics[width=0.4\textwidth]{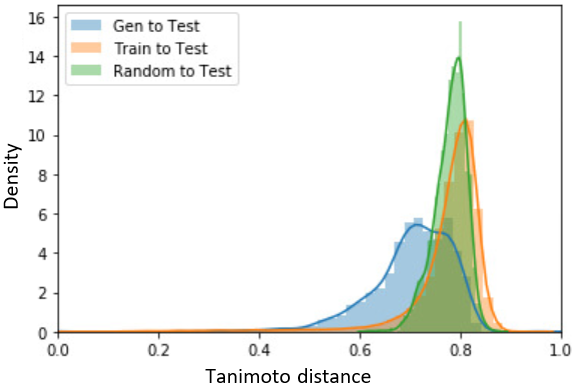}\label{fig:f1}}
  \hfill
  \subfloat[Similarity of the generated molecules for Tyr kinases to the training set. Several compounds show high similarity to training compounds. Whereas the training set lacks any Tyr kinase ligands it still contains compounds similar to those ligands.]{\includegraphics[width=0.4\textwidth]{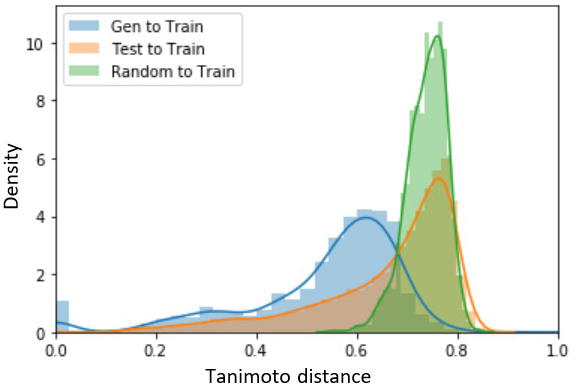}\label{fig:f2}}
    \hfill
  \subfloat[Similarity of the generated molecules for GPCRs to the test set that contains GPCR ligands.]{\includegraphics[width=0.4\textwidth]{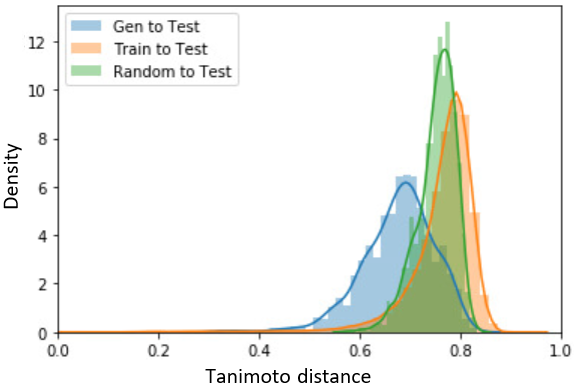}\label{fig:f3}}
    \hfill
  \subfloat[Similarity of the generated molecules for GPCRs to the training set that does not contain GPCR ligands.]{\includegraphics[width=0.4\textwidth]{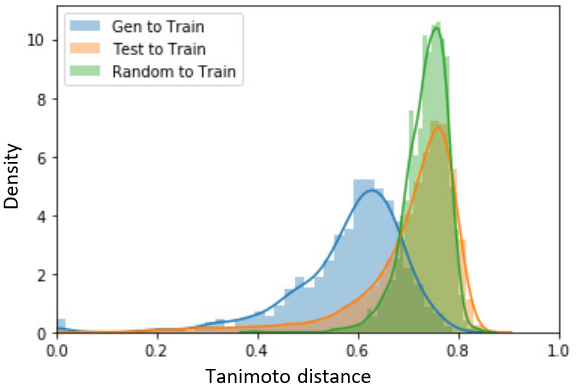}\label{fig:f4}}
  \caption{Similarity of the generated compounds for the Tyrosine kinase targets compared to the Tyrosine kinase test set (a) and the training set (b). The network generates more similar compounds for the Kinase targets than when compounds are selected randomly. In (b) it is observed that some compounds in the test set are similar to compounds of the training set. The reason for this are the existence of other kinase targets (non-Tyr kinases) in the training set that share similar compounds. The same plots are shown for the generation of GPCR ligand in (c) and (d). Density graphs were smoothed using kernel density estimation (KDE) available in the Seaborn library \citep{seaborn}.} \label{fig_similarities}
\end{figure}

\subsection{Encouraging compound diversity and novelty via reinforcement learning}
We generated 1653 unique novel molecules targeting 109 different Tyr kinases and 1672 compounds targeting 276 different GPCRs. 
Figure \ref{fig_similarities}a displays the similarity of our generated compounds compared to known Tyr kinase ligands (Test set). The generated compounds show overall a higher similarity to known Tyr kinase binders compared to the set of training molecules or randomly selected compounds from emolecules.  
Figure \ref{fig_similarities}b, however, shows that despite reinforcement learning for diversity, the generated compounds still show similarity with compounds from the training set that exceeds that of randomly selected molecules. 
There are, however, also several test compounds that have high similarity to at least one training compound. This observation is due to the fact that the training set contains other kinases (non-Tyrosine kinases) with ligands similar to Tyrosine kinase binders.

\begin{figure}[!htb]
  \centering
  \subfloat[Before reinforcement training]{\includegraphics[width=0.4\textwidth]{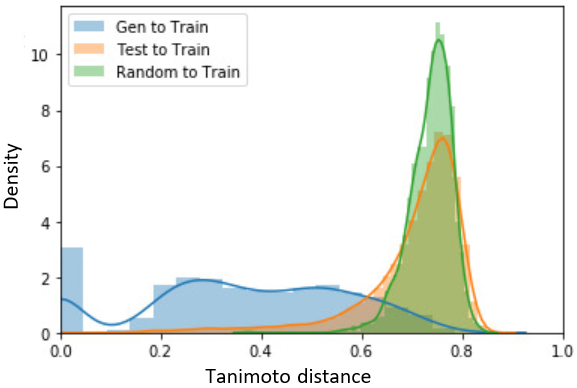}\label{fig:f1}}
  \hfill
  \subfloat[After reinforcement training]{\includegraphics[width=0.4\textwidth]{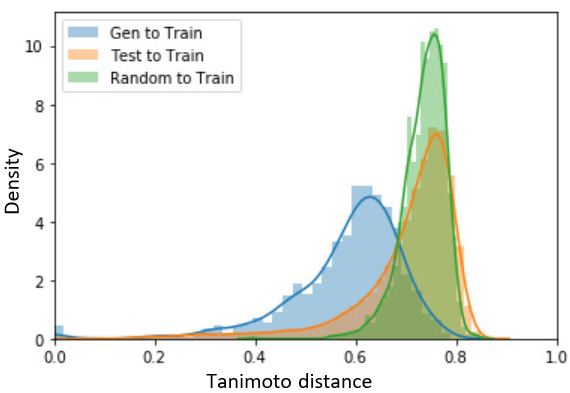}\label{fig:f2}}
  \caption{Encouraging diversity using reinforcement learning. Similarity distributions for GPCR targets (Tanimoto measure) are shown for the generated compounds before (a) and after (b) reinforcement training (blue color). It can be observed that without reinforcement the network generates mostly identical or very similar compounds to the training set. With reinforcement learning the similarity between generated and training molecules is significantly decreased.}
  \label{fig_with_no_RL}
\end{figure}

Similarly, compounds generated for GPCRs display a higher similarity to known GPCR ligand compared to training compounds or randomly selected molecules (Figure \ref{fig_similarities}c). While the generated compounds again have inherent similarities to some training set molecules, this similarity is less pronounced compared to the Tyrosine kinase case  (Figure \ref{fig_similarities}c), as the target family of GPCRs has no similar proteins in the training set.
Despite the remaining similarity between test and some training compounds, Figure \ref{fig_with_no_RL} demonstrates the effects of reinforcement learning to increase the novelty of the molecules. 

As discussed in the Material and Methods section, no member of the two target families used for the test of the models was present in the training set. Nevertheless, it is known that binding pockets of different protein families can share similarities in their binding pocket topology and properties. These similarities can results in sharing the same endogeneous ligand. For example, ATP can bind to many different target families \citep{atpproteome}. Another clinical observation of binding pocket similarity is the occurrence of side effects of drugs binding to multiple targets, primary or secondary. This fact can also be exploited in drug repurposing. 
Therefore, it is not unlikely that ligands similar to the training set are generated for the two target families of our test set.

\begin{figure}[!htb]
  \centering
  \subfloat[Tyrosine Kinase]{\includegraphics[width=0.7\textwidth]{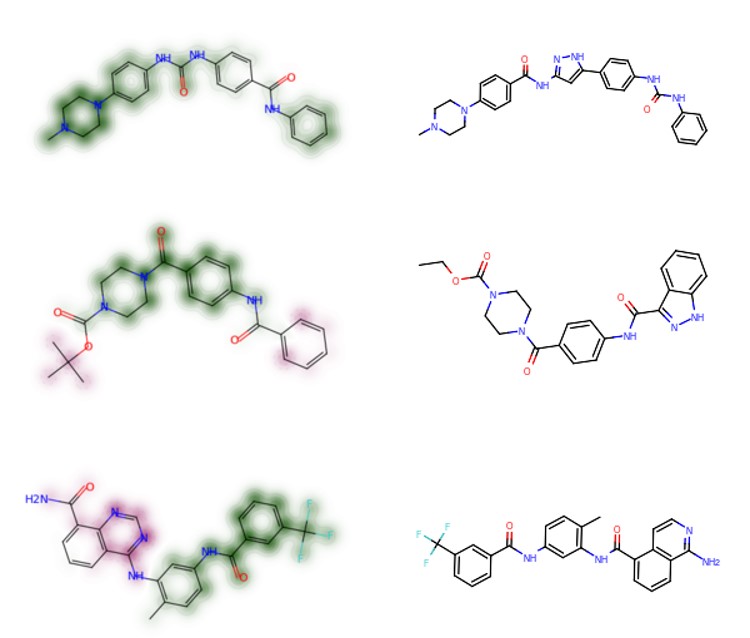}\label{fig:f51}}
  \vfill
  \subfloat[GPCR]{\includegraphics[width=0.7\textwidth]{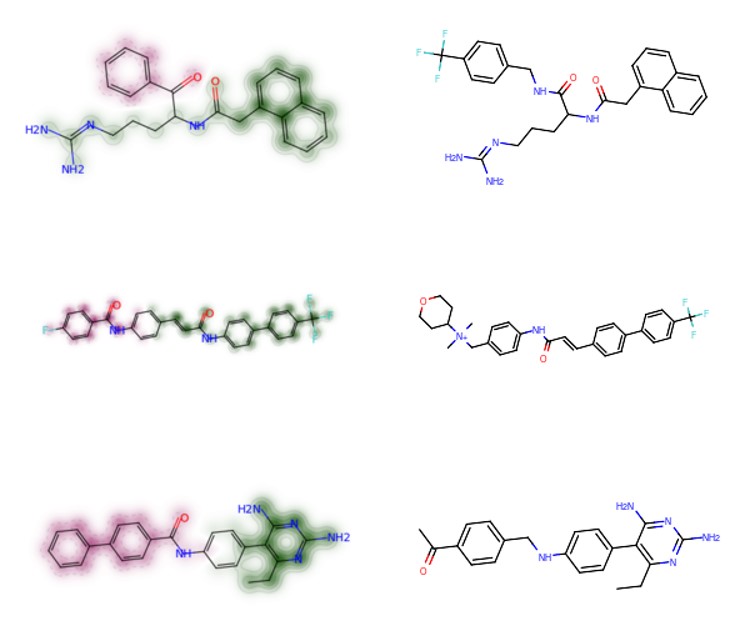}\label{fig:52}}

  \caption{Examples of 2D similarity maps of some generated compounds (left) with low Tanimoto distance to their most similar compounds from test sets (right). Substructures with high similarity are highlighted as green, dissimilar in red. Similarity maps were generated using RDKit's similarity map function \citep{similaritymaps}.}
  \label{fig_examples}
\end{figure}

Figure \ref{fig_examples} shows some examples of compounds generated for (a) Tyrosine kinases and (b) GPCRs together with their most similar known binder to the corresponding targets. They typically share a significant portion of the scaffold with known binders. 

\subsection{Comparison with benchmark models}
To compare the quality of our model, we used three baseline models, AAE, CharRNN, VAE as described in the MOSES framework \citep{moses}. 
Compounds were generated based on the same training set and evaluated against the same test sets as described above. The metrics used in MOSES and described in the Materials and Methods section were used to evaluate model performances (Table 1). Also the comparison of fingerprint similarity with the test set for the models was carried out (Figure 9). Whereas the three benchmark models generate compounds with similar fingerprint distribution to the training distribution, the compounds generated by our model are more similar to the target compounds forming the test set, despite the fact that those compounds or targets were never seen by the network (GPCR or Tyrosine Kinase ligands). 
This discrepancy between our and the benchmark models is encouranging but not surprising, since benchmark models produce compounds based on the probability distribution of tokens learned from the training data. On the other hand, in our model we bias compound generation by the protein's sequence. An embedding of the protein can be thought as a "barcode" that the token probabilities are conditioned on. 

Table 1 shows the comparisons of compound properties between the generated compounds by each model and the reference sets (Tyr Kinase and GPCR). The first three metrics show the similarity of the compounds to the reference set from different aspects. The first two metrics show the similarity of the sets in terms of fragments and scaffolds. Our model was able to generate compounds with smaller distance (higher similarity) in fragments for Tyrosine Kinase and GPCR targets and smaller distance in scaffold for GPCR targets, while scaffold distance is slightly higher for Seq2Mol model molecules. Distance to the nearest neighbor in test sets is lower for both Tyrisine Kinase and GPCR targets generated by our model. Benchmark models generated molecules with higher diversity. We believe this is due to the fact that benchmark models are not biased towards specific targets, so the diversity of their generated molecules is similar to the training set, which contains compounds for many different protein targets (therefore, more diverse in molecular structure). Table 2 shows metrics for basic molecular properties. It can be seen that the models were able to generate compounds with those properties within the acceptable range.

\begin{figure}[!htb]
  \centering
  \subfloat[Tyrosine Kinase]{\includegraphics[width=0.4\textwidth]{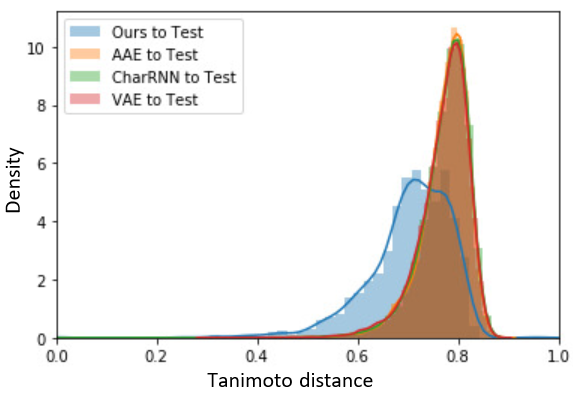}\label{benchmark:kinase}}
  \hfill
  \subfloat[GPCR]{\includegraphics[width=0.4\textwidth]{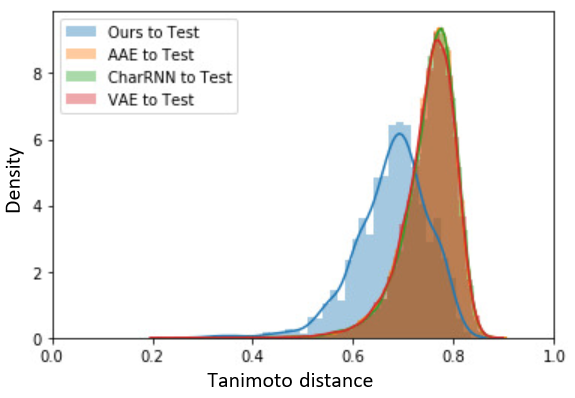}\label{benchmark:gpcr}}
  \caption{Comparison of similarity between generated compounds to the test sets for Tyrosine Kinase (a) and GPCR (b) using our model, AAE, CharRNN and VAE, respectively. While the benchmark models generate compounds with a similarity distribution similar to the training set, our model generates compounds more similar to the test sets, even though all models used the same set of compounds for training.}
\end{figure}
\afterpage{\FloatBarrier}

\subsection{Limitations}
\paragraph{Number of generated compounds} One limitation of the approach described in this manuscript is that the number of possible compounds that can be generated per target is limited to the maximum size of the beam width. One approach to enrich the pool of compounds is to use the generated compounds as seeds for other generative models to generate similar compounds with desired properties. 
\paragraph{Diversity versus relevance to the target} Another limitation of this approach is the ideal choice of the hyperparameters in reinforcement learning. The number of iterations and the amount of reward need to be carefully balanced. Otherwise the network may generate increasingly diverse compounds that are however no longer relevant to the target protein. 
In addition, pushing the model to generate highly diverse molecules might lead to the generation of compounds which are synthetically unfeasible.  Tuning the hyperparameters and therefore the amount of diversity to the training set may vary based on the needs of the drug design project and the target of interest.
\paragraph{Generation of ligands for targets with very similar sequence} There are cases that some targets can have very similar sequences (such as different isoforms of kinases). In such cases where the targets only differ in a few residues, the embeddings generated for the target proteins will be very similar, and the molecules generated for such targets will be highly similar. 
The reason is that the embedding of each residue over a sequence is generated and the embeddings are averaged to gain one vector representing the whole protein, therefore, the emebeddings for those residues does not affect the resulting vector of the protein.
In conclusion, the model will generate compounds relevant for a given target but will be unable to differentiate small variations in protein sequence and therefore protein-ligand interactions.
\paragraph{Targets with multiple binding sites} Some targets such as kinase proteins, have multiple binding sites and therefore different molecules with different structural properties may bind to the same target, although at different binding sites. 
For some targets, the binding site, the number of additional bindings sites or which binding site a specific compound binds to are unknown. This issue can create challenges for any drug design approach in which the 3D structure of the target, or the binding mode of the ligand is unknown. This is the case for a large portion of the experimental binding data in BindingDB. 
This problem could be overcome if only data for compounds with known binding site us used to train the network. In this case, it may also be possible only to use embeddings generated from the binding site sequences of the targets. The problem, however, is that currently no datasets large enough to train such a neural network are available.
\begin{table}[!htb]
  \label{metrics}
  \centering
    \caption{Various metrics measured using the MOSES framework for the generated compounds. Duplicates and invalid compounds were removed. Results for Tyrosine Kinase and GPCR targets are shown.}
\bigskip
\subfloat[Tyrosine Kinase]{
\begin{tabular}{lllll}
    \toprule
Metric & Seq2Mol & AAE & CharRNN & VAE \\
    \midrule

Fragment similarity distance & 0.847            & 0.929  & 0.942 & 0.941                                      \\
Scaffold similarity distance & 0.014            & 0.011       & 0.010 & 0.010                                   \\
Distance to the nearest neighbor & 0.566    & 0.662              & 0.662 & 0.659                           \\
Internal diversity & 0.765     & 0.852              & 0.856 & 0.855                            \\

\end{tabular}}

\subfloat[GPCR]{
\begin{tabular}{lllll}
    \toprule
Metric & Seq2Mol & AAE & CharRNN & VAE \\
    \midrule

Fragment similarity distance &  0.806  & 0.920 & 0.938  & 0.942                                 \\
Scaffold similarity distance & 0.017       & 0.057 & 0.062   & 0.055                         \\
Distance to the nearest neighbor    & 0.533              & 0.634 & 0.644 & 0.633                           \\
Internal diversity & 0.742     & 0.852 & 0.856 & 0.855   \\

\end{tabular}
}
\end{table}

\begin{table}[!htb]
  \label{metrics}
  \centering
    \small
    \setlength\tabcolsep{2pt}
    \caption{Various metrics measurements via MOSES framework for the generated compounds  after duplicates and invalid compounds are removed. Results for Tyrosine Kinase and GPCR targets are shown.}
\bigskip
\subfloat[Reference sets and our generated compounds corresponding to each set]{
\resizebox{\columnwidth}{!}{
\begin{tabular}{llllll}

    \toprule
Metric & Reference (Tyr Kinase) & Seq2Mol (Tyr Kinase) & Reference (GPCR) & Seq2Mol (GPCR) \\
    \midrule
LogP    & $3.99 \pm 1.46$ & $5.17 \pm 1.75$ & $4.50 \pm 1.57$ & $5.41 \pm 1.72$          \\
Synthetic accessibility        &   $2.93 \pm 0.55$ & $2.40 \pm 0.55$ & $3.06 \pm 0.68$ & $2.27 \pm 0.49$                            \\
  QED             & $0.47 \pm 0.17$ & $0.39 \pm 0.18$ & $0.48 \pm 0.19$ & $0.38 \pm 0.17$              \\
Natural product-likeness             & $-1.21 \pm 0.55$ & $-1.06 \pm 0.43$ & $-0.97 \pm 0.64$ & $-0.83 \pm 0.50$                                \\
  Molecular weight                & $450.27 \pm 85.20$ & $471.77 \pm 87.95$ & $465.57 \pm 95.74$ & $464.33 \pm 86.83$                              \\                           
\end{tabular}
}}

\subfloat[Benchmark models]{
\begin{tabular}{llllll}

    \toprule
Metric & AAE & CharRNN & VAE \\
    \midrule
LogP     & $2.46 \pm 1.00$ & $2.44 \pm 0.98$ & $2.47 \pm 0.94$            \\
Synthetic accessibility           & $2.49 \pm 0.47$ & $2.47 \pm 0.47$ & $2.43 \pm 0.46$                                 \\
 QED             & $0.80 \pm 0.10$ & $0.80 \pm 0.10$ & $0.81 \pm 0.09$                \\
Natural product-likeness            & $-1.65 \pm 0.59$ & $-1.68 \pm 0.64$ & $-1.67 \pm 0.64$                                 \\
  Molecular weight                & $318.89 \pm 30.91$ & $308.50 \pm 29.87$ & $304.64 \pm 28.65$                               \\
                                         
\end{tabular}
}

\end{table}

\section{Conclusion}
In this work we have developed a method for the de novo generation of molecules based on the sequence of the target. 
Unlike previous works in this area, our method does not need the knowledge of already known binders to a target protein as templates for molecule generation. The sequence of the target protein is sufficient to generate target-specific molecules instead. 
We showed that the pool of compounds generated for two large and important protein target familes, i.e. GPCRs and Tyrosine kinases, display meaningful similarity to  already known binders to these targets. 
With a continuous increase in number of experimentally resolved or computationally predicted protein structures, other types of protein embeddings based on 3D structure information may be used in the future, such as binding site embeddings based on 3D grids or graphs describing the binding site volume or arrangement of residues. respectively.













\bibliography{achemso-demo}

\end{document}